\documentstyle[preprint,aps]{revtex}
\preprint{\vbox{
\hbox{IFT-P.003/95 }
\hbox{DCR-TH 05/94 }
\hbox{IFUSP-P/1133 }
\hbox{hep-ph/9412331 }
\hbox{December 1994 }
}}
\begin{document}
\draft
\title{
Adiabatic matter effect with three generation neutrinos  \\
and the solar neutrino problem}
\author{M.M. Guzzo}
\address{
 Instituto de F\'\i sica Gleb Wataghin \\
Universidade Estadual de Campinas, UNICAMP\\
13083-970 -- Campinas, SP\\
Brazil}
\author{ O.L.G. Peres, V. Pleitez }
\address{ Instituto de F\'\i sica Te\'orica\\
Universidade Estadual Paulista\\
Rua Pamplona, 145\\
01405-900 -- S\~ao Paulo, SP\\
Brazil}
\author{R. Zukanovich Funchal }
\address{ Instituto de F\'\i sica da Universidade de S\~ao Paulo\\
01498-970 C.P. 20516 -- S\~ao Paulo, SP\\
Brazil}

\maketitle
\begin{abstract}
We find an exact analytic solution for the time evolution of a three Dirac
neutrino system adiabatically oscillating  in  matter,
 constructing explicitly  the relevant $3\times 3$ mixing matrix in matter.
Using this result we investigate the solar neutrino
data in a scenario where the neutrino masses are such that   $m_1\alt
m_2\ll m_3$, taking into account several phenomenological  constraints on
neutrino mixing angles and masses. A  solution of the solar neutrino
problem for large values of the  parameter $\delta m^2=m_2^2-m_1^2$ which are
not usually associated with a resonance is found. This is an essentially
three-generation effect.
\end{abstract}
\narrowtext

Most of the analyses of the neutrino oscillation
hypothesis  assume that this phenomenon involves only
two generations of neutrinos. It is difficult, however, to justify why
oscillations would not involve also the third family. From the
conceptual point of view, mixing and mass parameters required for
three generation oscillations are not different from mixing and mass
parameters that appear in this same phenomenon involving only two
generations of neutrinos. Why would such parameters related with the
third family vanish and not those ones leading to two neutrino
oscillations? In our opinion, the two generation analysis is just an
indicative approach to the more realistic three generation case.

The larger number of mixing and mass parameters in three neutrino oscillations
can be quoted as a difficulty to approach this scenario. There are three
mixing angles, one phase and two independent mass parameters that are,
in principle, free parameters. Due to this fact, when three generations
are considered in the literature \cite{three},
some assumptions have been made to restrict this parameter space
resulting that either only two family
transitions are effective or the parameters are fixed
arbitrarily.

In this letter
we find an exact analytic solution for the time evolution of a three Dirac
neutrino system adiabatically oscillating  in  matter,
constructing the three dimensional mixing matrix in matter, to  calculate the
electron
neutrino survival probability and compare it to the solar neutrino
data.

Assuming the
minimal extension of the
standard electroweak model when only three right-handed
neutrino singlets are introduced to generate Dirac neutrino masses,
a mixing can then occur among the three lepton flavors.
Therefore, neutrinos produced in weak processes are in general linear
combinations of the
mass eigenstates: $\nu_\alpha=\sum_iV_{\alpha i}\nu_i$ $(\alpha=e,\mu,\tau;~
i=1,2,3)$, where

\begin{equation}
V=\left(
\begin{array}{ccc}
c_\theta c_\beta \; \; \; & s_\theta c_\beta  \; \; & s_\beta \\
-s_\theta c_\gamma-c_\theta s_\gamma s_\beta \; \; \; & \; \; \;
c_\theta c_\gamma
-s_\theta s_\gamma s_\beta\; \; \;  & \; \; \;
s_\gamma c_\beta  \\
s_\theta s_\gamma -c_\theta c_\gamma s_\beta \; \; \; & \; \; \;
-c_\theta s_\gamma
-s_\theta c_\gamma s_\beta\; \; \;  & \; \; \;
c_\gamma c_\beta
\end{array}
\right).
\label{ckm}
\end{equation}

We have set to zero the CP violating phase in Eq.~(1).

The matter
effects~\cite{msw}
for the
generalized case of three generations are described by the time evolution
equation \cite{vb}
\begin{mathletters}
\label{se}
\begin{equation}
i\frac{d}{dt}\left(\begin{array}{c}
\nu_e \\ \nu_\mu \\ \nu_\tau\end{array}\right)=
\left[\frac{E_1+E_2}{2}{\bf1}+H\right] \left(\begin{array}{c}
\nu_e \\ \nu_\mu \\ \nu_\tau
\end{array}
\right)
\label{se1}
\end{equation}
where
\begin{equation}
H= V \left(
\begin{array}{ccc}
\frac{E_1-E_2}{2} & & \\ & -\frac{E_1-E_2}{2} & \\
& & E_3-\frac{E_1+E_2}{2}
\end{array}\right)V^{-1}+\left(
\begin{array}{ccc}
A & & \\
& 0 & \\
& & 0
\end{array}\right)
\label{h}
\end{equation}
\end{mathletters}
and $A=\sqrt2G_FN_e(t)$, with $G_F$ the Fermi constant, $N_e(t)$
the electron number density in the region reached by the neutrino at
the instant $t$.

In order to write the  neutrino survival probabilities when matter
effects are present, we would like to have the exact solution of the three
coupled differential equations given by Eq. (\ref{se}).
This has proven to be very difficult to obtain and
only approximate solutions~\cite{vb} has been achieved up to now.
In the following we will obtain  the explicit form of the relevant $3\times 3$
mixing matrix in matter. After that we will be able to  construct an exact
solution
for the three neutrino time evolution equations assuming that the
neutrino propagation is adiabatic everywhere.
In this case the problem is reduced to
diagonalize the matrix $H$ in Eq.~(\ref{se}). We
obtain the characteristic polynomial of $H$
\begin{equation}
\lambda^3+3a\lambda^2+3b\lambda+c=0,
\label{ce}
\end{equation}
where
\begin{equation}
3a=-Tr\,H,\quad
3b=H^m_{ee}+H^m_{\mu\mu}+H^m_{\tau\tau},\quad c=-det\, H
\label{defa}
\end{equation}
and $H^m_{\alpha\alpha}$ denotes the minor of the $H_{\alpha\alpha}$ elements.
We will not write explicitly the elements
$H_{\alpha\alpha'},\;\alpha,\alpha'=e,\mu,\tau$ since
they can  easily be obtained from Eqs.~(\ref{ckm}) and (\ref{se}).
The eigenvalues are
\begin{mathletters}
\label{eva}
\begin{equation}
\lambda_n=2\sqrt{-h}\cos\left(\frac{\Theta+2n\pi}{3}\right)-a,\;\;n=0,1,2 ;
\label{evaa}
\end{equation}
\begin{equation}
h=b-a^2,\quad g=c-3ab+2a^3,\quad
\cos\Theta=-\frac{g}{2\sqrt{-h^3}}.
\label{evab}
\end{equation}
\end{mathletters}
The respective eigenvectors are
\begin{equation}
\tilde V_{\alpha i}=\frac{\delta_{\alpha i}}{\delta_i},\quad
\alpha=e,\mu,\tau;\;i=1,2,3,
\label{eve}
\end{equation}
where
\begin{mathletters}
\label{def2}
\begin{equation}
\begin{array}{c}
\delta_{e1}=H_{e\mu}H_{\mu \tau}- H_{e\tau}(H_{\mu \mu}-\lambda_1), \quad
\delta_{\mu1}=H_{e\mu}H_{e\tau}-H_{\mu \tau}(H_{ee}-\lambda_1), \\
\delta_{\tau1}=(H_{ee}-\lambda_0)(H_{\mu\mu}-\lambda_1)-H_{e\mu}^2,
\quad
\delta_1=\left[\left(\delta_{e1}\right)^2  +
 \left(\delta_{\mu1}\right)^2  + \left(\delta_{\tau1}\right)^2
\right]^{\frac{1}{2}},
\end{array}
\label{def2a}
\end{equation}

\begin{equation}
\begin{array}{c}
\delta_{e2}=H_{e\mu}\delta_{\tau1}-H_{e\tau}\delta_{\mu1},\quad
\delta_{\mu2}=H_{e\tau}\delta_{e1}-(H_{ee}-\lambda_2)\delta_{\tau1}, \\
\delta_{\tau2}=(H_{ee}-\lambda_2)\delta_{\mu1}-H_{e\mu}\delta_{e1},
\quad \delta_2=\left[ \left(\delta_{e2}\right)^2  +
 \left(\delta_{\mu2}\right)^2  + \left(\delta_{\tau2}\right)^2
\right]^{\frac{1}{2}},
\end{array}
\label{def2b}
\end{equation}

\begin{equation}
\begin{array}{c}
\delta_{e3}=\delta_{\tau1}\delta_{\mu2}-\delta_{\tau2}\delta_{\mu1},\quad
\delta_{\mu3}=\delta_{\tau2}\delta_{e1}-\delta_{e2}\delta_{\tau1}, \\
\delta_{\tau3}=\delta_{\mu1}\delta_{e2}-\delta_{\mu2}\delta_{e1},\;
\quad \delta_3=\left[ \left(\delta_{e3}\right)^2  +
 \left(\delta_{\mu3}\right)^2  + \left(\delta_{\tau3}\right)^2
\right]^{\frac{1}{2}}.
\end{array}
\label{def2c}
\end{equation}
\end{mathletters}

Therefore the phenomenological eigenstates can be written in terms of the
matter
eigenstates
$\nu_\alpha=\sum_i
\tilde V_{\alpha i}\tilde\nu_i$, $(i=1,2,3)$, where the matrix $\tilde V$ can
be read from Eqs.~(\ref{eve}) and (\ref{def2}) and parametrize in terms of the
mixing angles in matter
as the matrix in Eq.~(\ref{ckm})  but
with $\theta\to\tilde\theta$, $\gamma\to\tilde\gamma$ and
$\beta\to\tilde\beta$. It is trivial now to
write down the averaged adiabatic survival probability of
finding a $\nu_\alpha$ produced at
the point $x_0$ inside the sun and detected at the point $x$ in the Earth's
surface
\begin{equation}
P_{\nu_\alpha\to\nu_\alpha}=\sum_i\vert\tilde{V}_{\alpha
i}(x_0)\vert^2\vert\tilde{V}_{\alpha i}(x)\vert^2.
\label{promat}
\end{equation}

The solar neutrino problem has been confirmed by many experiments.
In the following we will consider experimental data from
 Homestake(H), Kamiokande(K) and Gallex(G) experiments~\cite{exp1,fn1}.

Neutrinos produced in different reactions have different energies. While
$^7Be$ neutrinos are almost monochromatic~\cite{hx},
neutrinos produced in other source-reactions have different energy
spectra \cite{bu88} which have to be considered since, as we will see
in the following, the survival probability of the solar  neutrinos
is sensitive to their energy $E$ or their momentum $p$.
It is necessary also to take into account the value of the solar matter density
at the neutrino creation point $x_0$. We use the solar matter distribution
calculated through the Standard Solar Model  which is tabled in
Ref.~\cite{bu88}.
Notice that in Eq.~(\ref{promat}), since
 neutrinos are detected at the Earth's surface, the matrix elements at the
point $x$ are essentially those of the vacuum mixing matrix.

We can compare the theoretical neutrino flux ($\phi_{th}$) calculated
from the Standard Solar Model~\cite{bu88} with the observed flux
($\phi_{exp}$) measured by each experiment. The ratios
$R=\phi_{exp}/\phi_{th}$ are given by
$R(\mbox{H})=0.28\pm0.04$,
$R(\mbox{K})=0.49\pm 0.12$,and
$R(\mbox{G})=0.66\pm0.12$~\cite{exp1}.

Considering the neutrino  oscillations, these ratios can be calculated for
each experiment.
We take into account only the main source reactions of solar neutrinos which
are
sensible to each specific experiment.
For Homestake  we have
\begin{mathletters}
\label{pro}
\begin{equation}
R(\mbox{H})=0.78P^H(^8B)+0.14P^H(^7Be)+0.04P^H(^{15}O).
\label{proho}
\end{equation}
The neutrino flux measured by Kamiokande
facilities is not merely the electron neutrino one
since detector electrons will interact with other neutrino flavors via neutral
currents. For energies involved in the solar neutrino experiments, the
$\nu_e$-electron scattering cross section is about seven times larger than
other neutrino flavor ($\nu_\mu$-electron and $\nu_\tau$-electron) cross
sections. Hence, for Kamiokande, taking into account these neutral current
effects we have
\begin{equation}
R(\mbox{K})=P^K(^8B) + \frac{1}{7}[1-P^K(^8B)].
\label{proka}
\end{equation}
Finally, for Gallex
\begin{equation}
R(\mbox{G})=0.26P^G(^8B)+0.11P^G(^7Be)
+0.05P^G(^{15}O)+0.54P^G(pp).
\label{proga}
\end{equation}
\end{mathletters}
In Eqs.~(\ref{pro}) we use the notation
\begin{equation}
P^J(X)=\sum_{E>E^J_{thre}}f^X(E)P_{{\nu}_e\to
{\nu}_e}(E,\delta m^2,\theta,x,x_0).
\label{pes}
\end{equation}
$J=H,K$ and $G$ for Homestake, Kamiokande and Gallex; $X$ denotes
the particular source reaction of solar neutrinos and
$P_{{\nu}_\alpha\to{\nu}_\alpha}$ is given in Eq.~(\ref{promat}).
The threshold energy for each one of these experiments and
the energy spectrum of neutrinos produced in reaction $X$ are denoted
by $E^J_{thre}$ and $f^X(E)$, respectively. The spectra function $f^X(E)$
are given in Ref.~\cite{wh}.

As we said before, in Eq.~(\ref{promat}) there are still too many
free parameters: three vacuum mixing angles and three masses. Hence,
it is necessary to take into account the constraints from other
physical processes to fix some of the neutrino parameters before taking into
account the solar neutrino data.
The neutrino masses and mixing angles for the case of
$m_1\alt m_2\ll m_3$ have been determined in
Ref.~\cite{tau4} using $\tau$ leptonic decays, pion decays, $Z^0$
invisible width and end-point data from $\tau$ decay into five pions and
assuming
world average data for the ratio $G_\tau/G_\mu$.
Assuming the above mass hierarchy, the lower masses
$m_1,m_2$ and one angle $\theta$ remain undetermined, but
$m_3\sim165$ MeV, $11.54^o<\beta<12.82^o$ and $\gamma<4.05^o$.
Thus, we
have one mixing angle $\theta$ and two lightest neutrino mass difference
$\delta m^2=m_2^2-m_1^2$ to
be determined in neutrino oscillation processes.

Using Eqs.~(\ref{pro}) we have investigated the compatibility regions of the
three solar neutrino experiments. We have considered the region
$10^{-7}\, \mbox{eV}^2 \leq \delta m^2 \leq 10^{9}\,  \mbox{eV}^2$.
For $10^{-7}-10^{6} \, \mbox{eV}^2$ we have solution for either Homestake and
Kamiokande or Gallex and Kamiokande but not for all of them at the
same time.
The only two regions fitting the three experiments are shown in Fig. 1
at 95 \% C.L., we
have restricted the range of $\theta$ to be bellow $\pi$ as the survival
probability is symmetric in $\pi-\theta$. They correspond to:
$3.5\times 10^{6}\, \mbox{eV}^2 \leq \delta m^2 \leq 5\times10^{7}\,
\mbox{eV}^2$ with
$0.9 \leq \theta \leq 1.2$ radians and
$2\times 10^{6}\, \mbox{eV}^2 \leq \delta m^2 \leq 2\times 10^{7}\,
\mbox{eV}^2$ with
$1.4 \leq \theta \leq 2.2$ radians.
In fact we have found small regions of allowed values even at 68 \% C.L..

Some remarks are in order. With the parametrization of the mixing matrix
used in this work, the
resulting survival probability of the electron neutrino
is not sensible to the angle $\gamma$. Setting $\beta$, $m_3$, and $A$
equal to
zero, we recover the vacuum solution in two generation case~\cite{petcov}.
On the other hand
keeping $A \not = 0$ we recover the MSW solution in two generations.
In this case we have solution
for $\delta m^2 \approx 10^{-6} \, \mbox{eV}^2$ and
$\sin^2 2\theta \approx 1$, but only for two experiments at a time,
what is consistent with the fact that we are considering only the adiabatic
solutions~\cite{vignaud}.

General analytical descriptions  of three generation neutrino
oscillations are far from transparent.  We have found a solution to the
solar neutrino problem in a region of the relevant parameter space
which is not usually associated with resonances in the three neutrino
evolution equations \cite{vb}. This is a completely new feature compared
with the two neutrino MSW solution where only in the vicinity of a
resonance we can expect to find a dependence of the flavor survival
probability on the neutrino momenta. In that case, if we do not have a
resonance for the values of the oscillating parameters, all neutrinos
undergo the same survival probability and it is impossible to
conciliate all solar neutrino data.

Here we have looked for such dependence of the flavor survival
probability on the neutrino momenta investigating the behavior of the
relevant mixing angles in matter considering the momentum range of
neutrinos produced in the sun.  While pp-neutrinos have momenta not
larger than 0.44 MeV, $^8B$-neutrinos can present larger values of
momenta up to 15 MeV.  In Fig. 2 we show the results of this analysis.
We observed that for values of the mass difference $\delta m^2$ and the
vacuum mixing angle $\theta$ found to be relevant for the compatibility
of all solar neutrino data (see Fig. 1), a strong dependence of the
mixing angle in matter $\tilde\theta$ on the neutrino momentum $p$ is
present.

Summarizing. We have found an exact analytical
solution for the adiabatic transition probability in the three
generation neutrino case.  Applying our solution to an example where
one of the masses is rather large, of the order of several MeV, we have
shown that the mixing angles in matter strongly depend on the neutrino momentum
when we consider the  range of momenta physically interesting for the solar
neutrinos. Therefore a  solution to  the solar
neutrino problem can be achieved.

\acknowledgements

The authors would like to thank Funda\c c\~ao de Amparo \`a Pesquisa do
Estado de S\~ao Paulo (FAPESP) and
Con\-se\-lho Na\-cio\-nal de De\-sen\-vol\-vi\-men\-to Cien\-t\'\i
\-fi\-co e Tec\-no\-l\'o\-gi\-co (CNPq) for several financial supports.

\begin{figure}

\caption{The compatibility region of the three solar neutrino experiments
in the $\delta m^2 $-$\theta$ plane at $95\%$ C.L. .}
\label{f1}
\end{figure}

\begin{figure}
\caption{$\sin2\tilde\theta$ ($s_{2\tilde\theta}$) as a function of neutrino
momentum  $p$ for
$\delta m^2 = 1\times 10^7 \, \mbox{eV}^2$ and several values of
$\theta$.}
\label{f2}
\end{figure}

\begin{references}
\bibitem{fn1} SAGE  results are compatible with GALLEX ones.
\end{references}
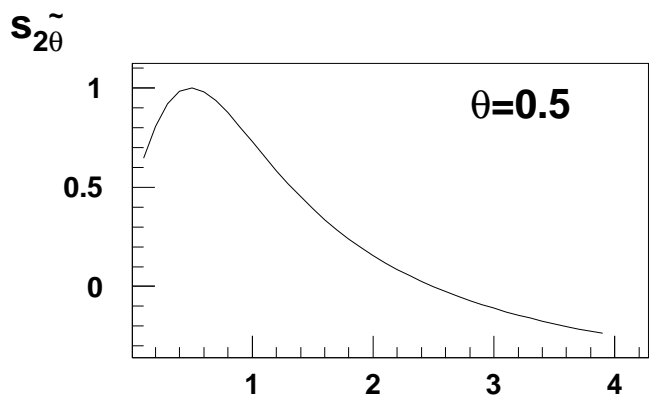
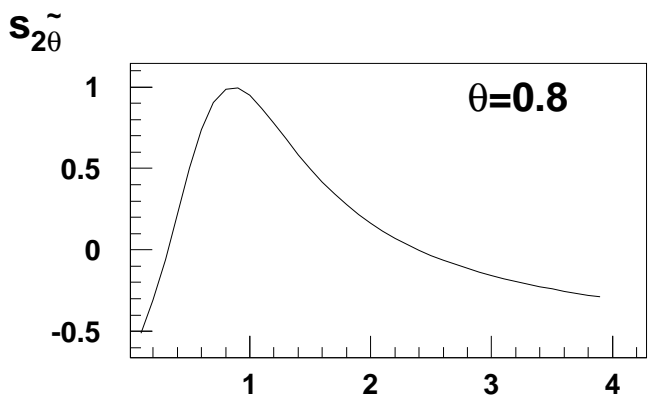
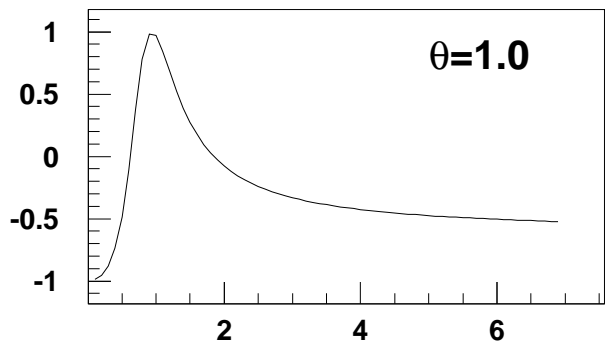
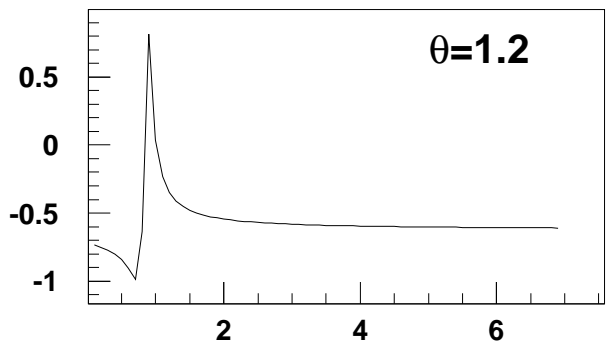
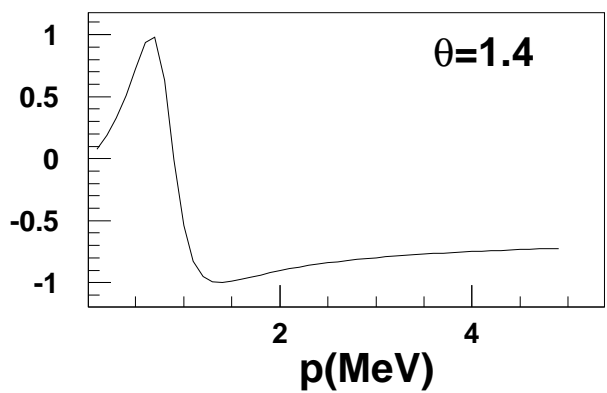
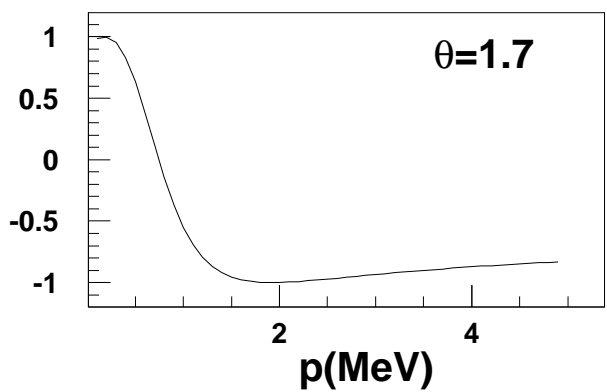
\end{document}